\documentclass[aps,prl,twocolumn,superscriptaddress]{revtex4-1}

\newcommand{\unit}[1]{\;\mathrm{#1}}

\usepackage{graphicx}

\begin{document}
\title{Direct Observation of Massless Domain Wall Dynamics in Nanostripes with Perpendicular Magnetic Anisotropy}
\author{J.~Vogel} \email{jan.vogel@grenoble.cnrs.fr}
\affiliation{Institut N\'{e}el, CNRS and UJF, BP166, 38042 Grenoble,
France}
\author{M.~Bonfim}
\affiliation{Institut N\'{e}el, CNRS and UJF, BP166, 38042 Grenoble,
France} \affiliation{Departamento de Engenharia El\'{e}trica, Universidade
do Paran\'{a}, CEP 81531-990, Curitiba, Brazil}
\author{N.~Rougemaille}
\affiliation{Institut N\'{e}el, CNRS and UJF, BP166, 38042 Grenoble,
France}
\author{O.~Boulle}
\affiliation{SPINTEC, UMR 8191, CEA/CNRS/UJF/Grenoble-INP, INAC, 38054
Grenoble Cedex, France}
\author{I.M.~Miron}
\affiliation{SPINTEC, UMR 8191, CEA/CNRS/UJF/Grenoble-INP, INAC, 38054
Grenoble Cedex, France}
\author{S.~Auffret}
\affiliation{SPINTEC, UMR 8191, CEA/CNRS/UJF/Grenoble-INP, INAC, 38054
Grenoble Cedex, France}
\author{B.~Rodmacq}
\affiliation{SPINTEC, UMR 8191, CEA/CNRS/UJF/Grenoble-INP, INAC, 38054
Grenoble Cedex, France}
\author{G.~Gaudin}
\affiliation{SPINTEC, UMR 8191, CEA/CNRS/UJF/Grenoble-INP, INAC, 38054
Grenoble Cedex, France}
\author{J.C.~Cezar}
\affiliation{ESRF, BP200, 38043 Grenoble, France}
\author{F.~Sirotti}
\affiliation{Synchrotron SOLEIL, L'Orme des Merisiers, Saint-Aubin,
91192 Gif-sur-Yvette, France}
\author{S.~Pizzini} \email{stefania.pizzini@grenoble.cnrs.fr}
\affiliation{Institut N\'{e}el, CNRS and UJF, BP166, 38042 Grenoble,
France}

\begin{abstract}
Domain wall motion induced by nanosecond current pulses in nanostripes with perpendicular magnetic anisotropy (Pt/Co/AlO$_x$) is shown to exhibit negligible inertia. Time-resolved magnetic microscopy during current pulses reveals that the domain walls start moving, with a constant speed, as soon as the current reaches a constant amplitude, and no or little motion takes place after the end of the pulse. The very low `mass' of these domain walls is attributed to the combination of their narrow width and high damping parameter $\alpha$. Such a small inertia should allow accurate control of domain wall motion, by tuning the duration and amplitude of the current pulses.
\end{abstract}

\maketitle

The interaction between conduction electron spins and local magnetization leads to a wealth of fascinating phenomena that have been extensively studied over the last fifteen years. This interaction allows, for instance, manipulating magnetic domain walls in nanostructures using current pulses \cite{Thomas2008,Boulle2011}. The temporal response of domain walls to the exciting current pulse is a key point for a better understanding of the interactions. It was recently shown that important transient effects can exist for domain walls in in-plane magnetized nanostripes, leading to a delayed domain wall motion with respect to the current pulse \cite{Chauleau2010,Thomas2010}. These transient effects, giving rise to domain wall `inertia' or an effective domain wall `mass' \cite{Saitoh2004}, are caused by deformations of the domain wall internal structure when a current or magnetic field is applied \cite{LiZhang2004,Thiaville2005,Thiaville2007}. Thomas \textit{et al.} \cite{Thomas2010} have shown that in the case of vortex domain walls these deformations can lead to a delay of several nanoseconds of the domain wall motion with respect to the current pulse and transient displacements of the order of 1.5~$\mu$m. Besides fundamental interest, such inertial effects potentially limit the use of domain walls in fast logic or memory devices, and transient times give an upper value of the excitation frequency. In this sense, domain walls with no inertia, i.e. that react instantaneously to an excitation, are highly desirable. Massless domain walls have been predicted theoretically in cylindrical magnetic nanowires with small diameter (below 50 nm) \cite{Yan2010}. The cylindrical symmetry should allow the magnetization direction inside the domain wall to rotate around the wire axis without changing the demagnetizing energy and without deformation of the domain wall structure. However, fabrication and experimental studies of such nanowires are difficult. In this paper, we will show that a very good approximation of massless domain walls can be obtained in more conventional magnetic nanostripes with perpendicular magnetic anisotropy and Bloch-type domain walls. Using time-resolved magnetic imaging, we show that in Pt/Co/AlO$_x$ nanostripes the domain walls move with constant velocity, without transient effects at both the beginning and the end of the pulses. We attribute this absence of inertial effects to the combination of a narrow domain wall width and a high damping parameter $\alpha$, leading to a large decrease of domain wall deformations with respect to N\'{e}el-type walls in nanostripes with a planar magnetization.

Current-induced domain wall motion (CIDM) has been studied by magnetic microscopy and electrical measurements \cite{Yamaguchi2004,Klaui2005,Adam2009,Pizzini2009,Uhlir2010,Hayashi2006,Tanigawa2009}, which have allowed determining the position and the internal structure of domain walls before and after the application of current pulses. Direct, microscopic observations of domain wall motion \textit{during} current pulses have been largely elusive until now. These observations are, however, essential for an unambiguous determination of inertial effects in domain wall motion. In this paper, we use time-resolved photoemission electron microscopy combined with x-ray magnetic circular dichroism (XMCD-PEEM) \cite{Vogel2003,Schonhense2006} to study current-induced motion of domain walls in Pt/Co/AlO$_x$ nanostripes with perpendicular anisotropy. The domain wall position was imaged during the application of 10-100 nanoseconds long current pulses. Our measurements show, in a direct way, that in these nanostripes the transient motion of the domain walls, which are expected to be of Bloch type in this material, is smaller than about 20 nm both at the beginning and at the end of the pulses, and thus much smaller than the displacements observed for N\'{e}el-type walls \cite{Chauleau2010,Thomas2010}.

Nanostructures based on asymmetric stacks of Pt/Co/AlO$_x$ are promising for new spintronic devices based on the manipulation of magnetization using current pulses \cite{Moore2008,Miron2010,Miron2011,Miron2011a}. The Rashba interaction, induced by the structural inversion asymmetry, leads to a high spin-torque efficiency and very high domain wall mobilities in this system \cite{Miron2011}.

\begin{figure}[ht!]
\includegraphics*[bb= 176 361 371 468]{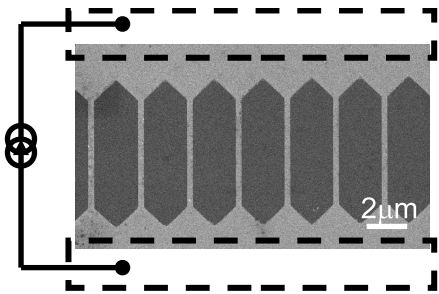}
\caption{\label{fig:SEM} Scanning Electron Microscopy image showing part of the twenty nanostripes connected in parallel to a pulsed voltage source. The Pt/Co/AlO$_x$ pads and lines show up in light grey, while the silicon substrate is dark grey. The gold contacts for current injection are schematically indicated with the dashed boxes.}
\end{figure}

Pt(3nm)/Co(0.6nm)/AlO$_x$ layers, deposited by magnetron sputtering on resistive Si, were patterned into twenty parallel 500 nm wide and 10 $\mu$m long stripes by e-beam lithography and Ion Beam Etching. Ti/Au electrical contacts were fabricated by UV lithography. A Scanning Electron Microscopy image of the sample is shown in Fig.~\ref{fig:SEM}. XMCD-PEEM measurements were performed at the TEMPO beamline of the SOLEIL synchrotron, using a Focus IS-PEEM. The magnetic contrast in the Co layer was optimized by subtracting two consecutive images collected at the Co L$_3$ absorption edge (778.1 eV) with 100\% left and right circularly polarized x-rays respectively. For each circular polarization, 60 images of 0.5 s were averaged, after correcting for possible image drifts. Temporal resolution was obtained by synchronizing the nanosecond current pulses applied to the sample with the x-ray pulses of the synchrotron single-bunch mode, where photon bunches reach the sample at a repetition rate of 846 kHz. The temporal evolution of the domain wall position in the nanostripes was obtained by recording images for different delays between the current and photon pulses \cite{Vogel2003,Uhlir2011}. If events are reproducible for each current pulse, the temporal resolution of this pump-probe technique is limited only by the duration of the photon pulses (about 50 ps) and the jitter between pump and probe (about 100 ps). The total acquisition time of 1 minute for each XMCD image implies that sequences of about $5 \times 10^7$ current (pump) and photon (probe) pulses were averaged.

In order to obtain a well-defined, reproducible initial domain wall position before each current pulse, the following procedure was used : i) starting from a saturated state, a 10 $\mu$s field pulse was applied perpendicular to the sample plane to create a domain wall in several nanostripes; ii) using a sequence of current pulses, the domain walls were driven to the top exit of the nanostripes, where the increasing cross section inhibits further motion in the up direction;  iii) bipolar current pulses with the same positive and negative amplitudes but a longer positive pulse were then applied for the stroboscopic measurements. The domain wall motion was studied during the negative part of the bipolar pulses, which drive the domain walls into the stripes, while the longer positive part of the pulses was used to reset the domain walls to their initial position. The measurements were carried out for several time delays, before, during and after the negative driving pulse, with time steps of 5-20 ns. Measurements were performed for current densities $J_1 = 7.7 \times 10^{11} \unit{A/m}^2$ and $J_2 = 1.3 \times 10^{12} \unit{A/m}^2$. The driving (reset) current pulse was 120 ns (160 ns) for J$_1$ and 30 ns (40 ns) for J$_2$. The risetime was of the order of 4 ns for all pulses.

\begin{figure}[ht!]
\includegraphics*[bb= 168 321 423 504]{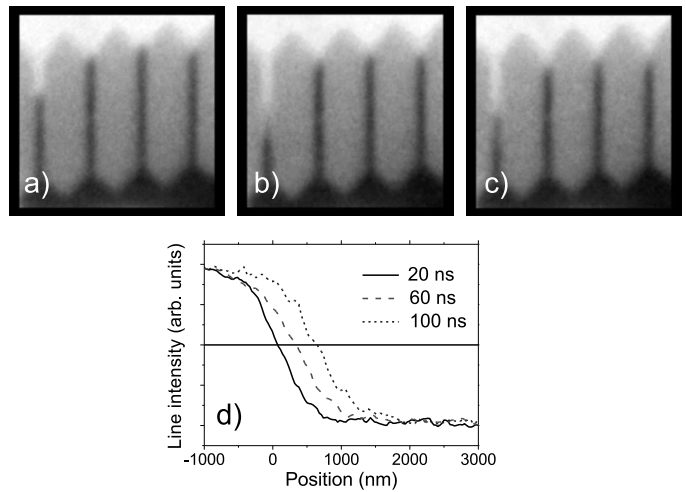}
\caption{\label{fig:images} Time-resolved XMCD-PEEM images taken during current pulses with a current density of $7.7 \times 10^{11} \unit{A/m}^2$, for delays of 20 ns (a), 60 ns (b) and 100 ns (c) after the beginning of the pulse. Line scans of the XMCD intensity averaged over the three stripes on the right are given in (d). The zero position in this graph corresponds to the initial position of the domain walls at the top entrance of the stripes.}
\end{figure}

In a PEEM microscope, the image of the sample is formed using the photoelectrons extracted from the sample surface. In order to obtain a sharp image, the potential between the focussing lens and the sample has to be adjusted accurately. Since the sample potential is modified during the current (voltage) pulses, upon scanning the delay between photon and current pulses it can be easily detected when they arrive on the sample at the same time, with an accuracy of about $100\unit{ps}$. During the current pulses a potential drop is present between the extremities of the stripes (about $3\unit{V}$ for the images of Fig.~\ref{fig:images}). This implies that the image cannot be well-focussed over the whole length of the stripes. This potential drop, and thus the image deformation, increases with increasing current density.

Figure~\ref{fig:images} shows XMCD-PEEM images for a selection of the twenty nanostripes present in our sample. The three images were taken during current pulses with a current density of $J_1 = 7.7 \times 10^{11} \unit{A/m}^2$, for delays of 20 ns, 60 ns and 100 ns after the onset of the drive current pulse. The images are corrected for the deformations induced by the voltage drop over the stripe length. In three of the four stripes shown in Fig.~\ref{fig:images} the domain wall positions are well-defined for each delay, and the domain walls move reproducibly from up to down \cite{movie} in the direction opposite to the electron flow \cite{Moore2009}. The behavior of the domain wall in the left stripe is much more stochastic, probably due to some strong pinning sites in the middle of the stripe, where the DW sometimes stays blocked for a certain time. This anomalous behavior was observed only in this stripe, while the dynamical behavior of the other lines was very similar, allowing data averaging to improve data quality.

\begin{figure}[ht!]
\includegraphics*[bb= 216 338 393 465]{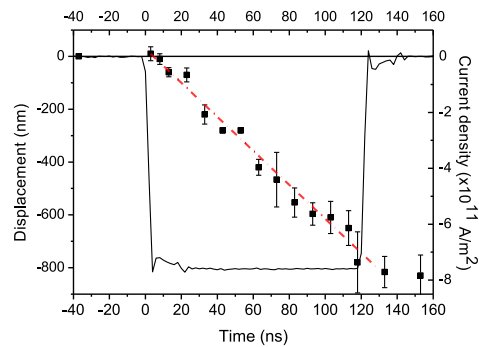}
\caption{\label{fig:displacements} (color online) The domain wall displacements, averaged over the three stripes on the right of the images in Fig.~\ref{fig:images}, as a function of time (black squares, left axis). The driving pulse with a current density of $7.7 \times 10^{11} \unit{A/m}^2$ is also shown (continuous line, right axis). The dashed line is a linear fit to the data points taken during the current pulse. The slope gives a domain wall velocity of 7 $\pm$ 1 m/s. The error bars are determined by the spread in displacement values for the three stripes, the error bar on the time is less than the width of the points.}
\end{figure}

The domain wall position for each time delay was obtained from the linescans of the XMCD intensity along each nanostripe. The domain wall displacement was defined with respect to the domain wall position before the driving pulse, corresponding to the top entrance of the stripes. The line scans averaged over the three rightmost stripes in Fig.~\ref{fig:images}(a)-(c) are given in Fig.~\ref{fig:images}(d). The corresponding domain wall displacements for all measured delays, with time steps of 5-10 ns, are given in Fig.~\ref{fig:displacements}. The driving current pulse is shown in the same figure.
The first important information provided by the time-resolved measurements is that the averaged domain wall displacement during the current pulse is linear in time, i.e. the DWs move at a constant average velocity. The velocity, obtained from a linear fit to the data, is 7 $\pm$ 1 m/s. Inertial effects should show up as a `delay' of the beginning of the constant velocity regime with respect to the beginning of the pulse \cite{Thomas2010}. The fit crosses $y = 0$ at $4.4 \pm 2.9 \unit{ns}$ from the beginning of the pulse. If we consider that the motion during the 4 ns risetime of the pulse is negligible, the motion starts thus within $0.4 \pm 2.9 \unit{ns}$ from the onset of the `plateau' of constant current density. The maximum time delay (including the error bar) of about $3 \unit{ns}$ corresponds to a maximum displacement delay of $3 \unit{ns} \times 7 \unit{m/s} = 21 \unit{nm}$ \cite{ChiralityReversal}.

We also measured the DW displacements for a 30 ns driving pulse with current density $J_2 = 1.3 \times 10^{12} \unit{A/m}^2$. Due to the higher voltage drop on the sample, the deformation of the images used to obtain these data was larger, leading to larger error bars. The average DW velocity obtained from the linear fit of the data points taken during the pulse is $45 \pm 10 \unit{m/s}$, with a delay with respect to the beginning of the current pulse of $0.8 \pm 5 \unit{ns}$. Transient effects are therefore small also for this higher current density.

In our previous work \cite{Miron2011}, we have shown that for current densities smaller than $2 \times 10^{12} \unit{A/m}^2$ the domain wall motion in Pt/Co/AlO$_x$ nanostripes is influenced by thermally activated depinning and can be described by the so-called creep law. Note that this thermally assisted depinning does not influence the domain wall speed averaged over a large number of displacements, but it leads to a distribution of domain wall positions that becomes wider when the average displacement is larger \cite{Moore2008}. This is reflected in the increase of the error bar as a function of time in Fig.~\ref{fig:displacements}. The domain velocities of $7 \pm 1\unit{m/s}$ for J$_1$ and $45 \pm 10\unit{m/s}$ for J$_2$ obtained here are in good agreement with the ones we obtained on similar samples with Kerr microscopy \cite{Moore2008,Miron2011}, where the average domain wall velocities were extracted from the slope of the displacement versus pulse duration.

Our data directly show that the transient motion of the domain walls is very small both at the onset and at the end of the current pulses, corresponding to a very small effective domain wall mass. According to Thiaville \textit{et al.} \cite{Thiaville2007}, the domain wall internal structure is modified under the action of a field or current pulse, and the transient displacement depends on the change of generalized angle, $\delta \phi$ : $\delta q = -\frac{\Delta}{\alpha}(1-\frac{\beta}{\alpha}) \delta \phi$. In this formula, $\delta q$ is the change in domain wall position, $\Delta$ the domain wall width at rest, $\alpha$ the damping parameter and $\beta$ the non-adiabatic parameter \cite{Thiaville2005,Zhang2004b}. For a Bloch wall in a nanostripe, $\phi$ corresponds to the tilt angle of the magnetization in the center of the wall with respect to the wall plane, in the direction parallel to the nanostripe. Above an angle $\phi$ of 45$^{\circ}$, the DW transforms into a N\'{e}el wall, then in a Bloch wall with opposite chirality (the so-called Walker limit). For the current densities used here, we have shown previously that the domain wall motion in our samples should be in the regime below the Walker breakdown \cite{Miron2011}, and the maximum value $\phi$ can take during domain wall propagation is thus 45$^{\circ}$.

One situation possibly leading to negligible transient effects ($\delta q \approx 0$) is when $\beta \approx \alpha$. However, previous experiments have shown that $\frac{\beta}{\alpha} \approx 2$ in our samples \cite{Miron2009}, with $\alpha \approx 0.5$ and $\beta \approx 1$, leading to $(1-\frac{\beta}{\alpha}) \approx -1$. The maximum transient displacement, for $\phi = 45^{\circ}$, would thus be $\frac{0.25\pi}{\alpha} \times 5\unit{nm} \approx 8 \unit{nm}$, where 5 nm is the approximate DW width $\Delta$ \cite{Miron2011}. This is in good agreement with our experiments. Since the transient motion is proportional to $\Delta$, it is expected to be smaller in systems with perpendicular anisotropy with Bloch-type domain walls, which are in general a factor ten to hundred narrower than N\'{e}el-type domain walls in in-plane systems. However, the small transient motion in our system is favored also by the relatively large value of $\alpha$ : for a damping parameter $\alpha=0.02$ (typical for permalloy) and $\frac{\beta}{\alpha} = 2$, the maximum transient motion would be about 200 nm, much larger than our experimental observation. Finally, the angle $\phi$ and thus the transient motion should be reduced by the s-d mediated Rashba field \cite{Miron2011}.

In conclusion, we have used time-resolved magnetic imaging to directly reveal the absence of transient effects during current-induced domain wall motion in Pt/Co/AlO$_x$ nanostripes with perpendicular anisotropy. We attribute the negligible domain wall mass to the combination of a narrow domain wall width, a large value of the damping parameter $\alpha$ and the s-d mediated Rashba field transverse to the nanostripes. Added to the large domain wall velocities obtained in these systems \cite{Miron2011} and the good reproducibility of the domain wall displacements, such a small domain wall inertia should allow a fast and accurate control of domain wall motion at high repetition rate, by tuning the duration, frequency and amplitude of the current pulses. Our measurements also show the extreme robustness of these Pt/Co/AlO$_x$ nanostripes, since many billions of pulses with current densities higher than $1 \times 10^{12} \unit{A/m}^2$ could be applied at high frequencies without changing their physical properties.

We acknowledge the invaluable technical and experimental help of E.~Wagner, D.~Lepoittevin, L.~Delrey, Z.~Ishaque and A.~Hrabec, as well as O.~Fruchart for discussions. Nanofabrication was performed at the Institut N\'{e}el/CNRS `Nanofab' facility in Grenoble. This work was partially supported by the Agence National de la Recherche through projects ANR-07-NANO-034 `Voice' and ANR-11-BS10-008 `Esperado', and by the Fondation Nanosciences.

\bibliographystyle{apsrev}


\begin{thebibliography}{26}
\bibitem {ChiralityReversal} Note that the current direction of our driving pulse
    is opposite to the one in the reset pulse, meaning that the current direction
    reversed at the beginning of eah driving pulse. In Ref. [21] it was shown that
    a reversal of the current direction is associated to a change of chirality of
    the Bloch domain wall, since the s-d mediated Rashba field switches direction.
    For a current density of $1.4 \times 10^{12} \unit{A/m}^2$, a delay in motion of 20 nm
    was observed, corresponding to a delay time of 0.2 ns at the corresponding
    DW velocity of 100 m/s. This delay time is well within the error bar of about
    3 ns in the present experiment.
\end{thebibliography}

\end{document}